\documentclass[letter]{aa}  

\usepackage{graphicx}
\usepackage{txfonts}
\begin{document}

   \title{The fate of planetary cores in giant and ice-giant planets}

   \subtitle{}

   \author{
     S. Mazevet
          \inst{1,2}
          \and
          R. Musella\inst{1}
     \and
     F. Guyot\inst{3}
         }

   \institute{Laboratoire Univers et Th\'eories, Universit\'e Paris Diderot, Observatoire de Paris, PSL University, 5 Place Jules Janssen, 92195 Meudon France.\\
     \email{stephane.mazevet@obspm.fr}
     \and
     CEA-DAM-DIF, 91280 Bruy\`eres le Ch\^atel, France.\\
     \and
     Institut de Min\'eralogie de physique des Mat\'eriaux et de Cosmochimie (IMPMC), Museum National d'Histoire Naturelle, Sorbonne Universit\'e, IRD, CNRS, Paris, France
             }

   \date{\today}

 
  \abstract
      {The Juno probe that currently orbits Jupiter measures its gravitational moments with great accuracy. Preliminary results suggest that the core
        of the planet  may be eroded. While great attention has been paid to the material properties of elements constituting
        the envelope, little is known about those that constitute the core. This situation clutters our interpretation the Juno data and modeling of giant planets and exoplanets in general}
      {We calculate the high-pressure melting temperatures of three potential components of the cores of giant planets, water,
        iron, and a simple silicate, MgSiO3, to investigate the state of the deep inner core.}
      {We used {\sl \textup{ab initio}} molecular dynamics simulations  to calculate the high-pressure
        melting temperatures of the three potential core components.
        The planetary adiabats were obtained by solving the hydrostatic equations
        in a three-layer model adjusted to reproduce the measured gravitational moments. Recently developed {\sl \textup{ab initio}}
        equations of state were used for the envelope and the core. 
      }
      {We find that the cores of the giant and ice-giant planets of the solar system differ because the pressure-temperature
        conditions encountered in each object correspond to different regions of the phase diagrams. For Jupiter and Saturn, the results are compatible with a diffuse
        core and mixing of a significant fraction of metallic elements in the envelope, leading to a convective and/or a double-diffusion regime. We also find that their
        solid cores vary in nature and size throughout the lifetimes of these planets. The solid cores of the two giant
        planets are not primordial and nucleate and grow as the planets cool.
        We estimate that the solid core of Jupiter is 3 Gyr old and that of Saturn is 1.5 Gyr old. The situation is less extreme for Uranus and Neptune, whose cores are only partially melted.
      }
      { To model Jupiter, the time
        evolution of the interior structure of the giant planets and exoplanets in general, their luminosity, and the evolution
        of the tidal effects over their lifetimes, the core should be considered as crystallizing and growing rather than gradually mixing into the envelope due to the solubility of its components.}
        


   \maketitle
%

   \section{Introduction}
   The core-accretion model, which assumes that giant planets form by accretion of hydrogen and helium around a solid
   core, is a reference model in planetary modeling. It is used to explain the rapid formation of large giant
   planets of several hundred Earth masses by the rapid runaway accretion of gaseous hydrogen-helium material from
   the planetary nebula if the core size is beyond a critical size \citep{Pollack_etal96}. In conjunction
   with basic hydrodynamics arguments, it brings a simplified picture of the  interior structures of these planets
   as two adiabatic layers of varying densities in hydrostatic equilibrium, one for the hydrogen-helium envelope and a second corresponding to
   a primordial core that is enriched in heavy elements. This simple picture carries over to evolutionary models
   where the time variation of the luminosity is obtained by integrating the energy dissipated by
   these layered structures backward in time.  The composition of these structures is often assumed to be fixed during the planet lifetime \citep{guillot_1995,guillot_2015}.

   The measurements of gravitational moments of the giant planets of the solar system, their luminosity, atmospheric
   composition, and the mass-radius relationships obtained for the thousands of exoplanets that are now detected
   provide mounting evidence that this model of planetary interiors needs to be improved \citep{Baraffe_2014,heller_2018}.
   For the giant planets of the solar system, the gravitational
   moments measured for either Jupiter or Saturn are not compatible with a pure and homogeneous hydrogen-helium envelope \citep{Nettelmann_2012,Nettelmann_2013}. This suggests
   that in addition to a varying helium concentration, metallic elements such as water or silicates may be present deep in the envelope and close to the core.
   This view of the Jupiter interior was re-enforced by a recent analysis of the Juno measurements, where
   a diffuse core was invoked to explain the measured gravitational moments \citep{Wahl_etal17,Debras_2019}.

   The latest formation models, which are more in line with global simulations of the formation of planetary systems, further
   suggest that late accretion of planetesimals may also explain the varying metallicity that is observed in the atmospheres of
   giant planets \citep{Lin_2007,Alibert_2018}. While this approach also leads to a non-uniform density in the envelope, the fate of the primordial
   core during the planet lifetime currently remains a major unknown in deciding how the underlying hypothesis
   of the core accretion model might be relaxed. This is by extension also true of the two- or three-layer models. This mostly stems from the lack of
   accurate physical properties for the elements that may constitute the core at the extreme pressure-temperature
   conditions that are encountered deep within the envelope and at the core-envelope boundary \citep{Baraffe_2014}.

   A first step to address
   this open question was recently obtained by considering the miscibility of potential core constituents within a pure
   hydrogen plasma \citep{Wilson_2012,Wilson_2014,Soubiran_2015}.  This assumes that a solid core is slowly
   dissolving into the envelope as the planet cools down. These calculations further consider that miscibility for a given element in a pure hydrogen
   plasma is representative of the miscibility in a planetary envelope consisting of a hydrogen-helium mixture with impurities. This approach consequently neglects the effect
   of potential additional elements that are also dissolved in the envelope in a significant fraction that may even become dominating as we approach the core. To proceed beyond this step
   becomes rapidly untractable because the miscibility of all the potential constituents in varying concentrations in the core and envelope needs
   to be considered.  We here suggest that considering the high-pressure melting properties
   of potential elements that constitute the core is an alternative way to build quantitative interior structure models that are more consistent with their cooling history.

   \section{High-pressure melting properties of potential core materials}

   We considered three basic components that might constitute the primordial core: a simple silicate, MgSiO$_3$,
   water, H$_2$O, and iron, Fe. We included iron as a possible core component because dynamical simulations indicate that migration
   of the planet below the ice line may cause a non-negligible amount of iron to accumulate that in turn may finally rest   within the core.   

\begin{figure}
   \centering
   \includegraphics[width=8cm,height=6cm]{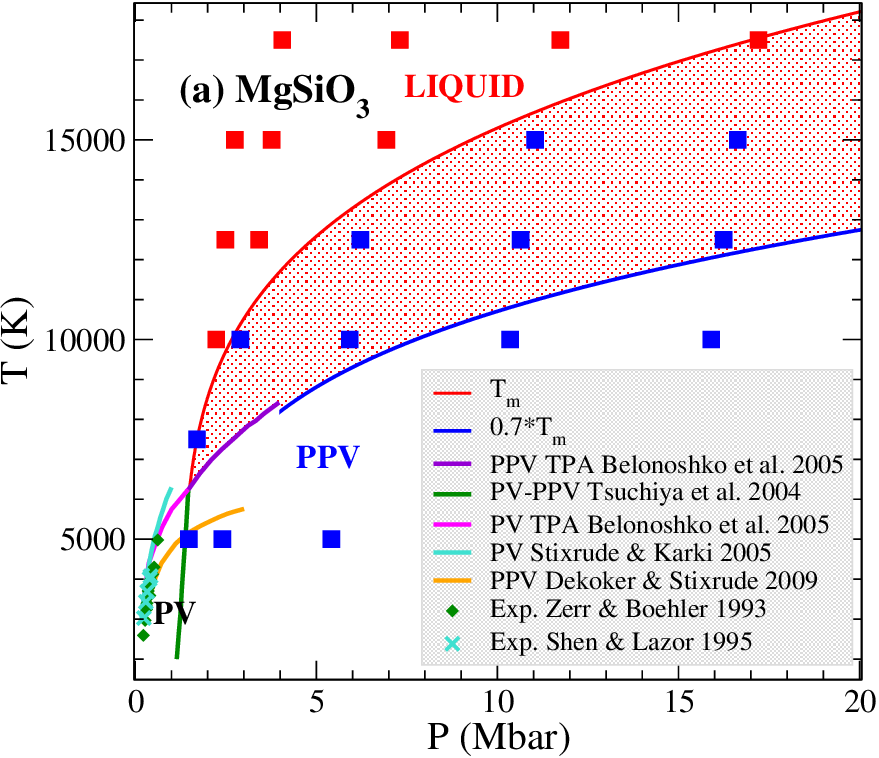}
   \includegraphics[width=8cm,height=6cm]{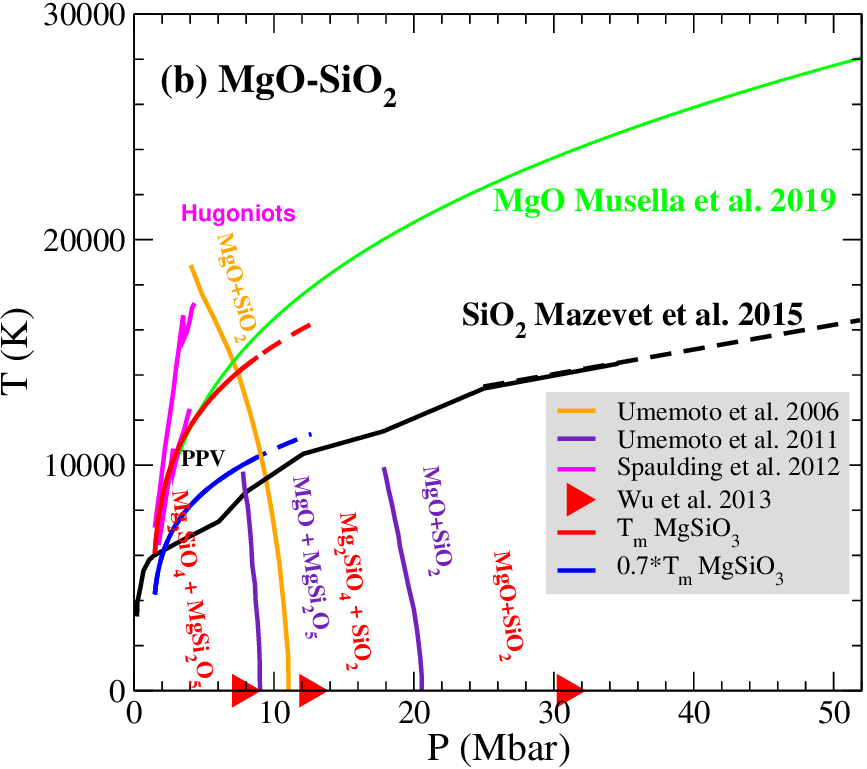}
   \caption{(a)  High-pressure melting temperature obtained for the PPV phase of MgSiO$_3$. Conditions where the simulations
     equilibrate in the liquid or solid states are indicated as red and  blue squares, respectively. Previous theoretical results for the PV and PPV
     melting temperatures are indicated in the legend. Two-phase results are indicated as TPA. Experimental data in the PV phase are indicated as (Exp.).
     The shaded area represents the conditions where melting can occur. (b) Dissociation pathways predicted for MgSiO$_3$ beyond $10~$Mbar and high-pressure
     melting temperatures for the dissociation products SiO$_2$ and MgO.}
   \label{fig1}
\end{figure}

   We first turn to the high-pressure properties of MgSiO$_3$ , which presents many polymorphs at low pressures.
   Because we are mainly concerned with the conditions that are encountered in the core of giant planets, we   focus on the post-perovskite (PPV) phase of MgSiO$_3$ , which is considered as the stable one in the $1-10~$Mbar range. Figure \ref{fig1}-a shows the
   equation-of-state (EOS) points calculated using molecular dynamics simulations (see the Appendix). We obtained the pressure dependence of the melting
   temperature we calculated, $T_m$, by adjusting the semi-empirical Simon law \citep{Poirier} to the {\sl \textup{ab initio}} results (see the Appendix).
   The melting temperature steadily increases with pressure and reaches $18,000~$K at $20~$Mbar. The liquid or solid states were obtained by considering the mean square
   displacement once the simulation was equilibrated. When we compare our results with previous estimates that are available up to $4~$Mbar \citep{belonoshko_2005,stixrude_2009}, we see that this approach leads to a significant overestimation. This overheating effect is well known. It comes from simulations that are performed at fixed
   volume while using a limited number of particles. The melting can be overestimated by up to 30\%. Because more refined approaches are beyond reach for this system,
   we applied a conservative coefficient of 0.7 to our estimate (noted 0.7$\ast T_m$ in figure \ref{fig1}-a). Figure \ref{fig1}-a shows that this
   allows us to extend the results obtained by \citet{belonoshko_2005} while at the same time
satisfying   the pressure dependence found in our simulations. Our results further rule out the predictions of \citet{stixrude_2009}.
 
   Direct inspection of the stress tensor obtained from the simulations also shows that the off-diagonal components become
   non-negligible for pressures beyond $10~$Mbar. This indicates that the PPV phase becomes unstable beyond this pressure range. This
   result is in line with previous work \citep{umemoto_2006} that found no stable post-PPV phase for MgSiO$_3$ and
   dissociation into simpler compounds beyond $10~$Mbar. The current understanding of the MgSiO$_3$ phase diagram is displayed in figure
   \ref{fig1}-b. Since the pioneer work of \citet{umemoto_2006}, the dissociation pathway has been refined to include intermediate compounds found in the $10-30~$Mbar
   range. This includes Mg$_2$SiO$_4$ and MgSi$_2$O$_5$ \citep{umemoto_2011,Wu_2013}. Because the details of this dissociation pathway
   are likely a second-order effect for planetary modeling, we approximated the melting temperature of MgSiO$_3$
   as the combination of the high-pressure melting temperature obtained from our simulations,
   corrected for overheating to match the results of \citet{belonoshko_2005} at $4~$Mbar, with the high-pressure melting temperature of
   the simpler components, MgO and SiO$_2$, beyond $20~$Mbar that we obtained previously \citep{mazevet_2015,musella_2019}.

\begin{figure}
   \includegraphics[width=8cm,height=6cm]{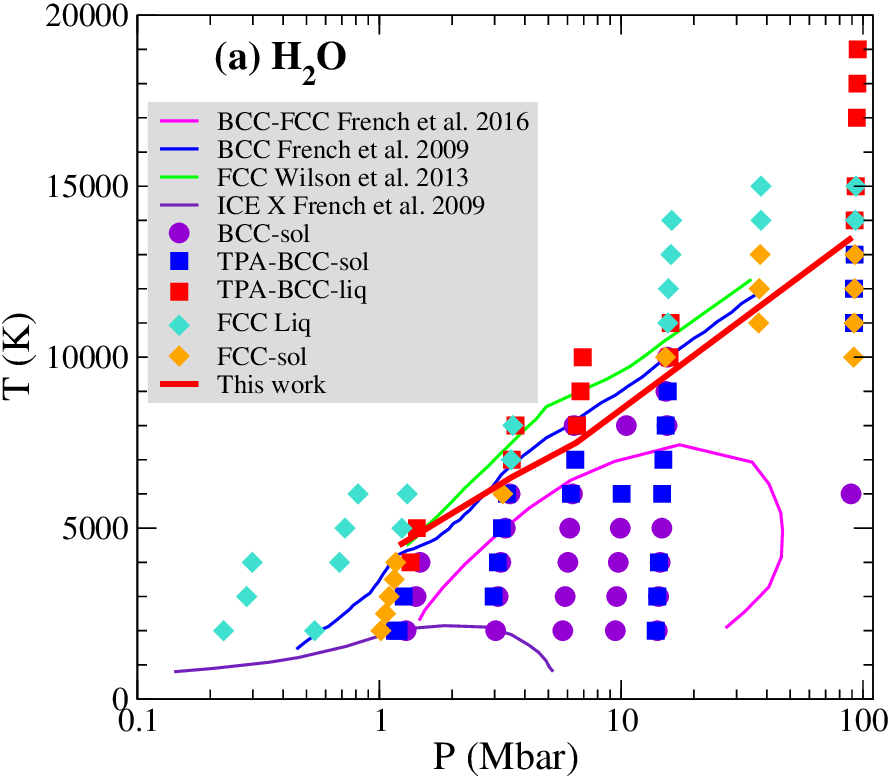}
   \includegraphics[width=8cm,height=6cm]{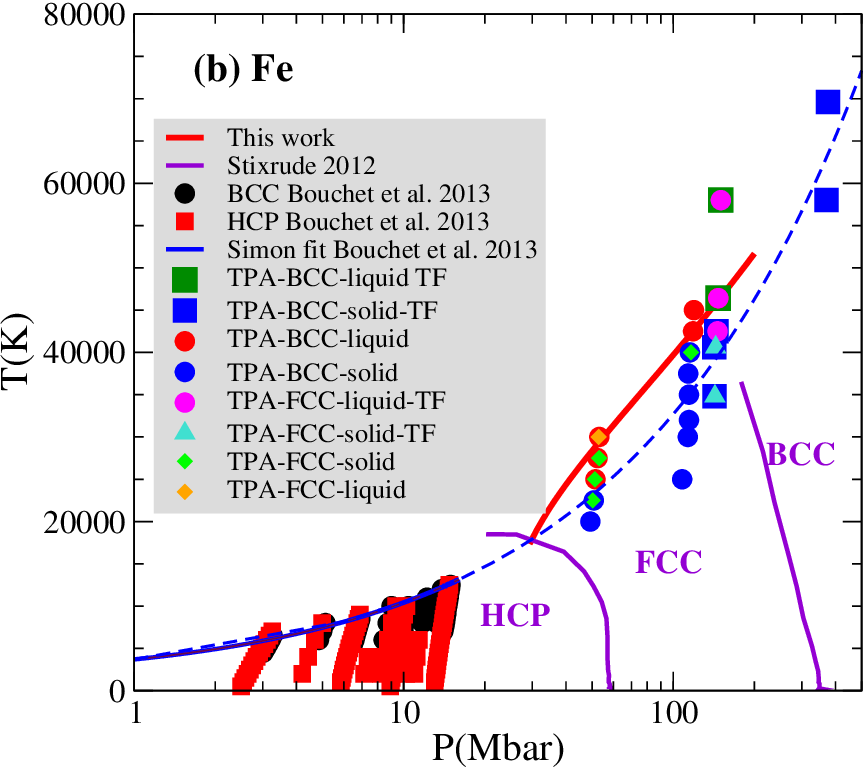}
   \caption{(a) Comparison of the high-pressure melting temperature of super-ionic water in the FCC and BCC phases obtained in
     previous work. The simulations results are indicated as symbols. The red line represents the stability region of super-ionic
     water obtained in this work. (b) Comparison between the high-pressure melting temperature of iron calculated in previous
     work. Simulation results are indicated as symbols corresponding to different phases, and the stability regions of the different phases are indicated in the figure with the boundaries calculated by \cite{stixrude_2012}.  }
   \label{fig2}
\end{figure}

   In figure \ref{fig2}-a we show the results obtained for dense water. This work complements previous studies
   performed by \cite{french_2009} and \cite{wilson_2013}, who reported for the super-ionic phase that
   the oxygen atoms either lie in FCC or BCC structures.
   Because different melting temperatures were obtained, we first revisited these calculations by performing
   direct melting simulations and considered both structures as initial states (see the Appendix). Figure \ref{fig2}-a shows that our estimate of the BCC
   structure agrees well with the result of \citet{french_2009}.
   In contrast with the findings of \citet{wilson_2013}, we find that the FCC structure is not more stable than the BCC structure above $1~$Mbar because the high-pressure melting
   temperature in the FCC phase is equal to or lower than in the BCC phase. Like \cite{french_2016}, we also find that the FCC structure appears more stable around $1~$Mbar
   and below this pressure.

   This shows that the stability of the BCC structure is a good estimate for planetary modeling.
   We also find that the BCC structure is unstable just below melting, which indicates that
   another crystalline structure may be present under these conditions. To further investigate whether super-heating is an issue for the
   super-ionic state of dense water, we performed two-phase simulations in the BCC structure. We find that simulation of direct melting
   tends to overestimate the stability of the super-ionic phase by a few thousand Kelvin. This result is consistent with
   the fact that the volume change is small through this transition \citep{french_2016}. Figure \ref{fig2}-a shows our
   new estimate of the pressure-temperature domain where the super-ionic state should be considered. This reaches $13,500~$K
   at $100~$Mbar and includes the FCC phase below $1~$Mbar.     

   We finally turn to the behavior of the last potential element that may be present in planetary embryos: iron.
   To estimate the stability of the solid phase of iron in the pressure range considered here,
   we extended our previous calculations \citep{morard_2011,bouchet} up to $100~$Mbar (see the Appendix).  We further performed
   two-phase simulations by considering the stable phases of iron as predicted by linear response theory \citep{stixrude_2012}. Figure \ref{fig2}-b
   shows that solid iron is found in an HCP state up to the $30-60~$Mbar range. The FCC crystalline structure is predicted to be the most
   stable one beyond this pressure and up to $200~$Mbar. Beyond $200~$Mbar, the BCC structure is predicted to be the most stable \citep{stixrude_2012}.

   Our two-phase simulations confirm this overall result. At $60~$Mbar, the high-pressure melting
   temperature of the FCC phase is higher than that of BCC by $5000~$K, indicating that it is the most
   stable phase when both temperature and anharmonic effects are taken into account. We further point out that we used
   fewer atoms in the simulation cell in the BCC than in the FCC phase.  We previously found at lower pressures
   that this can lead to an underestimation of the melting temperature by a few thousand Kelvin. It is thus likely that the difference
   between the BCC and FCC melting temperatures reported here is slightly smaller.  
 
   Beyond $100~$Mbar, we switched to the computationally more efficient Thomas Fermi molecular dynamics approach to converge in the
   number of particles used.  Probably due to the influence of the $3$s state, 
   figure \ref{fig2}-b shows that the Thomas Fermi regime is not yet fully reached at $100~$Mbar because pressures in the BCC phase are overestimated by 15\%. Despite this limitation,
   the method allows for an additional estimate of T$_m$.  Figure \ref{fig2}-b shows that this method
   predicts that the BCC phase is more stable with T$_m$ up to $3000~$K higher than for the FCC phase. This tends to confirm the results obtained using linear response
   theory.  The {\sl \textup{ab initio}} results obtained with a smaller
   number of atoms in the BCC phase do not confirm this finding and tend to suggest that the FCC phase may still be the stable phase. 
   Resolving this question is beyond the scope of this paper, therefore we considered
the FCC melting temperature from $30$ to
   $100~$Mbar and the Thomas Fermi result beyond $100~$Mbar, as indicated
in figure\ref{fig2}-b, to extend the pure iron melting temperature beyond $15~$Mbar.
   
   \section{Implications for giant planet cores}

\begin{figure}
     \includegraphics[width=8cm,height=6cm]{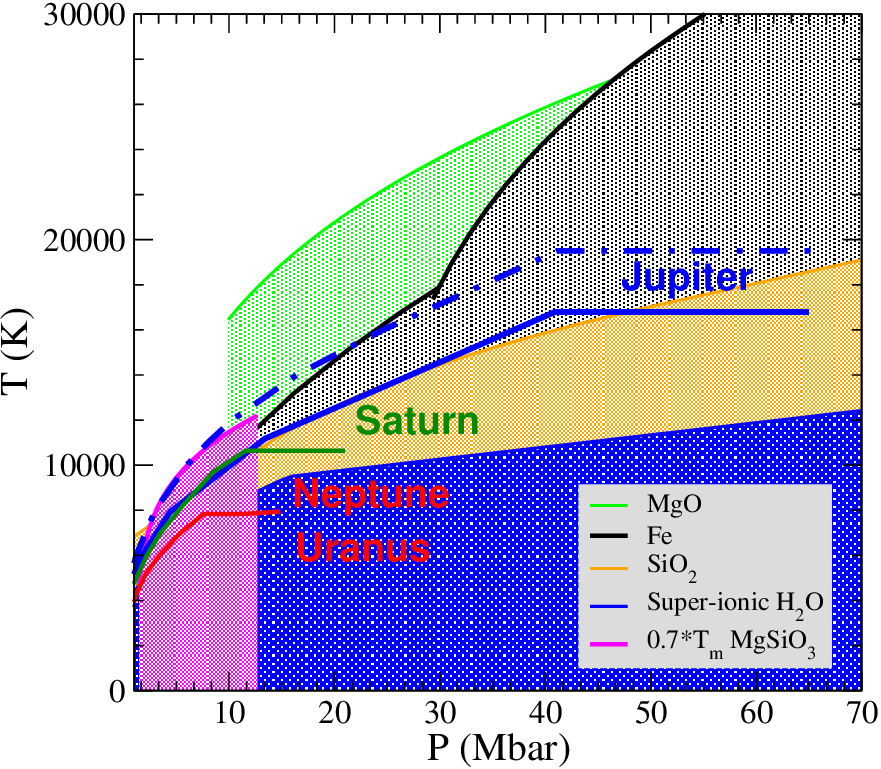}
     \caption{Comparison between the high-pressure melting temperatures we obtained and density-temperature profiles for the planets
       of the solar system (red lines). The dashed regions correspond to P-T conditions where the constituent is in a solid state. The adiabats for Saturn, Neptune, and Uranus are from \citet{guillot_2015}. For Jupiter, we show the adiabats obtained by \citet{Nettelmann_2012} (dashes) and \citet{Militzer_2013} (solid). }
     \label{fig3}
   \end{figure}

   We now use these high-pressure melting temperatures we calculated to estimate the state of the core in giant and ice-giant planets.
   Figure \ref{fig3} shows the temperature profile obtained for Jupiter when two different models based on different
   {\sl \textup{ab initio}} EOSs \citep{Nettelmann_2012,Militzer_2013} for the envelope and the core were used. The origin of this difference
   has been well documented elsewhere \citep{Miguel_2016,mazevet_2019a}. We concentrated on the region corresponding to the core. This
   corresponds to a plateau in the temperature profile because the core is described by an isothermal profile. 
   When we consider Jupiter, the comparison between the predicted interior profiles and the melting temperatures calculated
   previously indicates that if they are present within the core, MgO and Fe are clearly in a solid state.
   Figure \ref{fig3} also shows that H$_2$O is not in a super-ionic state. H$_2$O can thus potentially convect and mix with the envelope and/or establish a regime of
   layered convection with a significant density gradient due to gravitational settling \citep{leconte}.     
  
   Figure \ref{fig3} also shows that the SiO$_2$ melting temperature is in between the predictions of
   \citet{Nettelmann_2012} and \citet{Militzer_2013} for the interior profile of Jupiter. This indicates that depending on the interior model
   that is used, SiO$_2$ should either be completely melted up to the center or be present with a solid fraction close to the core center.
   In both cases, figure \ref{fig3} shows that SiO$_2$ is melted at the core-envelope interface and can also mix into the envelope and/or participate
   in a layered convection regime.
   Considering the low-temperature profile, figure \ref{fig3} shows that the current state
   of the Jupiter core should be considered as made of solid MgO, possibly including some solid iron, surrounded by solid SiO$_2$,
   some fraction of SiO$_2$ present at the time of formation potentially mixing into the envelope, and H$2$O, which can fully mix within the envelope.
   We point out that this picture provides a stronger physical basis to the latest interpretation of the JUNO data, where a partially
   diluted core was invoked to reproduce the gravitational moments to high order \citep{Wahl_2017,Debras_2019}. This result is obtained
   by only considering the melting temperature of possible core components and does not rely on their solubility.

   For Saturn, our calculations suggest a core containing
   solid MgSiO$_3$ together with iron that is either dissolved in silicate or is present as metal.  The high-pressure melting temperature predicted from our
   simulations is however rather close to the isotherm calculated for the core by \citet{guillot_2015}. The difference between the
   two is only about $2000~$K. This points to uncertainties on the presence of solid MgSiO$_3$ in the core of Saturn. The situation is clearer for
   the two ice giants Uranus and Neptune. We find that MgSiO$_3$ is expected to be in a solid state in their cores.  The calculated core conditions are in this case
   well below the high-pressure melting temperatures. The situation differs for super-ionic water, for which we find that the core conditions for the two ice-giants
   reported by \citet{guillot_2015} lie very close to the melting temperature. It is thus likely that water is also partially mixed into the envelope of ice-giant planets.
   Overall, the calculations of the melting temperatures performed in this work clearly suggest that the cores of the giant and ice-giant planets of the solar system are currently partially
   mixed with the envelope. They also suggest that a double-diffusive or layered convection regime involving a mixture of the various elements that constitute the initial
   core is likely taking place for the giant planets and that it does not have a clear interface between a hydrogen-helium envelope and a solid core. To further explore this assumption, we
   now turn to the time evolution of the cores of the two giant planets as they cool down.
\begin{figure}
   \includegraphics[width=8cm,height=6cm]{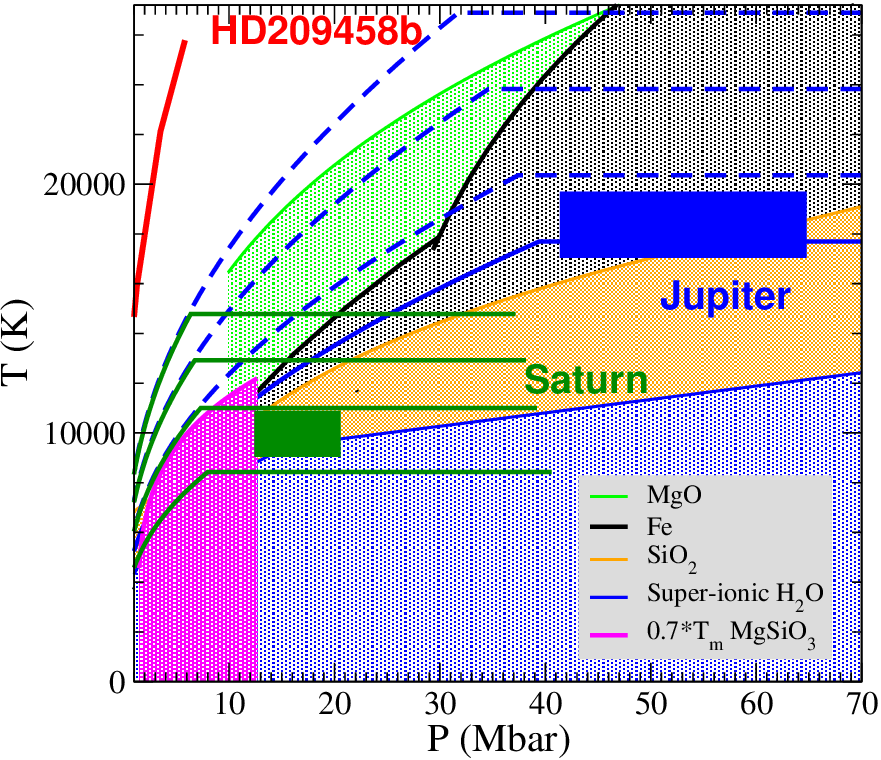}  
   \caption{ Comparison between the high-pressure melting temperatures obtained with density-temperature profiles of Jupiter (blue lines) and
     Saturn (green lines) where the surface temperatures at 1 bar, $T_{atm}$, are increased from current conditions to $300$K. The hotter adiabats correspond to higher surface
     temperatures. The rectangles indicate the conditions for the core we showed in figure \ref{fig3}.}
   \label{fig4}
\end{figure}

   We show in figure \ref{fig4} adiabats for Jupiter and Saturn that we obtained by increasing the atmospheric temperature $T_{atm}$ at $1$bar.
   The calculations were performed using our newly developed {\sl \textup{ab initio}} EOSs for hydrogen,
   helium, water, and MgO \citep{chabrier_2019,mazevet_2019b,musella_2019}. We used a three-layer model that consists
   of an H-He envelope, and a core made of H$_2$O, and MgO or MgSiO$_3$ for Jupiter and Saturn, respectively. We solved the standard
   hydrostatic equations and optimized the fraction of each layer using the theory of figures to the third order to match the
   radius and the first three gravitational moments. By considering $T_{atm}$
   measured by the Galileo probe at a pressure $P_{atm}=1~$bar and $T_{atm}=165$K \citep{galileo}, we obtain a profile for Jupiter in which the temperature of the core is slightly higher than the temperature obtained by \citet{Militzer_2013}, that is, favoring a low-temperature profile
   for Jupiter. Additional details can be found in \citep{mazevet_2019a}.

   Figure \ref{fig4} shows the interior profiles obtained for Jupiter by increasing $T_{atm}$ from current conditions to $200~$K, $250~$K,
   and $300~$K while keeping all the other parameters fixed. At $T_{atm}=200~$K, SiO$_2$
   is completely melted up to the center of the core. At temperature $T_{atm}$ greater than $250~$K, figure \ref{fig4} indicates that 
   the MgO core starts to be melted at the core-envelope interface. The MgO core is no longer completely solid
   and can mix into the envelope or participate in a double-diffusive or layered convection regime. We assume here
   that viscosity of the solid at these extreme conditions prevents it from participating in a convective or layered convective mode. 

   The situation is similar for Saturn. We first point out that at $T_{atm}=140~$K, the measured temperature at $1~$bar, our
   three-layer model adjusted to reproduce the gravitational moments predicts a core temperature  $2000~$K lower than the calculations
   of \citet{guillot_2015}. This comes from the different H-He EOSs that were used and translates for the case of Saturn the sensitivity to the H-He EOSs
   found for Jupiter \citep{mazevet_2019a}. Figure \ref{fig4} shows that under
   current conditions, the core of Saturn contains solid MgSiO$_3$. At the core-envelope boundary, H$_2$O is in a
   liquid state and can thus mix into the envelope or enter a double-diffusive or layered convection regime. A fraction of H$_2$O is in a super-ionic state deeper within the center
   of the core. This suggests that the core is currently made of super-ionic water and MgSiO$_3$ that is probably dissociated
   into solid MgO and solid SiO$_2$ deep within the core. When $T_{atm}$ is increased to $200~$K, MgSiO$_3$ is no longer in a
   solid state at the core-envelope boundary, while water can no longer be found in a super-ionic state. As $T_{atm}$ is increased to $250~$K and $300~$K,
   figure \ref{fig4} shows that the size of the solid core reduces as the core isotherm crosses the MgSiO$_3$ high-pressure melting curve at successively
   higher temperatures. 
   
   Using standard evolution models for the giant planets \citep{guillot_1995}, we can directly relate the interior profiles
   calculated at different $T_{atm}$ to earlier states in the history of the planets. Using the time evolution of $T_{atm}$ calculated in
   \citet{guillot_1995}, we estimate that $T_{atm}=300~$K corresponds to Jupiter at $1~$Gyr after formation, that is, $3.5~$Gyr ago, while $T_{atm}=250~$K corresponds to the
   state of Saturn $1.5~$Gyr ago. Because $T_{atm}$ further increases earlier on for both planets, this indicates that
   the cores were likely completely melted during the early stages after formation, in the same way as for HD209458b, for example. This suggests that soon after formation, the cores
   of both planets were completely melted and likely entered a double-diffusive or layered convection regime involving all the
   primary constituents of the core, with a likely significant density gradient due to gravitational settling.
   Figure \ref{fig4} also shows that Jupiter and Saturn not only have different cores at the present time, but also followed different evolutionary tracks. As the planets cooled down,
   MgO first precipitated in the core of Jupiter and no longer contributed to a double-diffusive or layered convection regime, but SiO$_2$ is only recently doing so. Conversely,
the core of   Saturn likely formed more recently, experienced first precipitation of MgSiO$_3$, and is experiencing super-ionic H$_2$O precipitating at the present time.

   \section{Discussions and outlook}
   {\  Our calculations of the high-pressure melting temperatures of several components that might be present in the core of giant
   planets indicate that the solid cores of these objects are not primordial. After an initial regime where all the constituents likely convected and/or entered a double-diffusive or layered
   convection regime, the solid fraction evolved throughout the lifetime of the planet with a time-dependent solidification of its components. Although less spectacular, a similar situation
   with an only partially melted core is suggested for the ice giants. Our calculations
   thus bring a different perspective to the diffuse core approximation that has been invoked to reproduce the gravitational moments of Jupiter that were measured  by the Juno probe. They further suggest
   that the solubility of the various core constituents in a pure hydrogen plasma may not be relevant. In effect, no clear interface between the core and
   the H-He interface exists throughout the lifetime of the planets of the solar system, with water acting as a buffer between the core forming and the H-He envelope.
   We further point out that if melted, layered convection or double-diffusive convection of the core constituents can take place, which might affect the
   energetics and the evolutionary track of each planet differently.  We suggest that this may be particularly the case for Saturn, for which current models
   struggle to explain the current luminosity, and for giant exoplanets in general. Our calculations indicate that the
   time evolution of their luminosity should consistently take the evolving state of the core into account. Finally, our results suggest that the varying nature
   of planetary cores should also be considered when tidal effects of the giant planets of the solar system on their satellites are estimated.}

   \section{Acknowledgment}
    Part of this work was supported by the ANR grant PLANETLAB 12-BS04-0015 and from Paris Sciences et Lettres (PSL) 
    university through the project origins and conditions for the emergence of life. This work was performed using HPC
    resources from GENCI- TGCC (Grant 2019- A0030406113)
   \bibliographystyle{aa} 
   \bibliography{gcore} 

\begin{thebibliography}{41}
\expandafter\ifx\csname natexlab\endcsname\relax\def\natexlab#1{#1}\fi

\bibitem[{{Alibert} {et~al.}(2018){Alibert}, {Venturini}, {Helled}, {Ataiee},
  {Burn}, {Senecal}, {Benz}, {Mayer}, {Mordasini}, \& {Quanz}}]{Alibert_2018}
{Alibert}, Y., {Venturini}, J., {Helled}, R., {et~al.} 2018, Nature Astronomy,
  2, 873

\bibitem[{{Allen}~M \& {Tidsley}(1989)}]{tidsley}
{Allen}~M, P. \& {Tidsley}, D.~J. 1989, {Computer Simulations of Liquids}
  (Oxford Science Publications)

\bibitem[{{Baraffe} {et~al.}(2014){Baraffe}, {Chabrier}, {Fortney}, \&
  {Sotin}}]{Baraffe_2014}
{Baraffe}, I., {Chabrier}, G., {Fortney}, J., \& {Sotin}, C. 2014, in
  Protostars and Planets VI, ed. H.~{Beuther}, R.~S. {Klessen}, C.~P.
  {Dullemond}, \& T.~{Henning}, 763

\bibitem[{{Belonoshko} {et~al.}(2005){Belonoshko}, {Skorodumova}, {Rosengren},
  {Ahuja}, {Johansson}, {Burakovsky}, \& {Preston}}]{belonoshko_2005}
{Belonoshko}, A.~B., {Skorodumova}, N.~V., {Rosengren}, A., {et~al.} 2005,
  \prl, 94, 195701

\bibitem[{{Bouchet} {et~al.}(2013){Bouchet}, {Mazevet}, {Morard}, {Guyot}, \&
  {Musella}}]{bouchet}
{Bouchet}, J., {Mazevet}, S., {Morard}, G., {Guyot}, F., \& {Musella}, R. 2013,
  Physical Review B, 87

\bibitem[{{Chabrier} {et~al.}(2019){Chabrier}, {Mazevet}, \&
  {Soubiran}}]{chabrier_2019}
{Chabrier}, G., {Mazevet}, S., \& {Soubiran}, F. 2019, \apj, 872, 51

\bibitem[{{de Koker} \& {Stixrude}(2009)}]{stixrude_2009}
{de Koker}, N. \& {Stixrude}, L. 2009, Geophysical Journal International, 178,
  162

\bibitem[{{Debras} \& {Chabrier}(2019)}]{Debras_2019}
{Debras}, F. \& {Chabrier}, G. 2019, \apj, 872, 100

\bibitem[{{French} {et~al.}(2016){French}, {Desjarlais}, \&
  {Redmer}}]{french_2016}
{French}, M., {Desjarlais}, M.~P., \& {Redmer}, R. 2016, \pre, 93

\bibitem[{{French} {et~al.}(2009){French}, {Mattsson}, {Nettelmann}, \&
  {Redmer}}]{french_2009}
{French}, M., {Mattsson}, T.~R., {Nettelmann}, N., \& {Redmer}, R. 2009,
  Physical Review B, 79

\bibitem[{{Gonz{\'a}lez-Cataldo} {et~al.}(2014){Gonz{\'a}lez-Cataldo},
  {Wilson}, \& {Militzer}}]{Wilson_2014}
{Gonz{\'a}lez-Cataldo}, F., {Wilson}, H.~F., \& {Militzer}, B. 2014, \apj, 787,
  79

\bibitem[{{Gonze} {et~al.}(2009){Gonze}, {Amadon}, {Anglade}, {Beuken},
  {Bottin}, {Boulanger}, {Bruneval}, {Caliste}, {Caracas}, {C{\^o}t{\'e}},
  {Deutsch}, {Genovese}, {Ghosez}, {Giantomassi}, {Goedecker}, {Hamann},
  {Hermet}, {Jollet}, {Jomard}, {Leroux}, {Mancini}, {Mazevet}, {Oliveira},
  {Onida}, {Pouillon}, {Rangel}, {Rignanese}, {Sangalli}, {Shaltaf}, {Torrent},
  {Verstraete}, {Zerah}, \& {Zwanziger}}]{2009CoPhC.180.2582G}
{Gonze}, X., {Amadon}, B., {Anglade}, P.~M., {et~al.} 2009, Computer Physics
  Communications, 180, 2582

\bibitem[{{Guillot} {et~al.}(1995){Guillot}, {Chabrier}, {Gautier}, \&
  {Morel}}]{guillot_1995}
{Guillot}, T., {Chabrier}, G., {Gautier}, D., \& {Morel}, P. 1995, \apj, 450,
  463

\bibitem[{{Guillot} \& {Gautier}(2014)}]{guillot_2015}
{Guillot}, T. \& {Gautier}, D. 2014, arXiv e-prints, arXiv:1405.3752

\bibitem[{{Helled} \& {Guillot}(2018)}]{heller_2018}
{Helled}, R. \& {Guillot}, T. 2018, {Internal Structure of Giant and Icy
  Planets: Importance of Heavy Elements and Mixing} (Handbook of Exoplanets,
  ISBN 978-3-319-55332-0. Springer International Publishing AG, part of
  Springer Nature, 2018, id.44), 44

\bibitem[{{Jollet} {et~al.}(2014){Jollet}, {Torrent}, \& {Holzwarth}}]{jollet}
{Jollet}, F., {Torrent}, M., \& {Holzwarth}, N. 2014, Computer Physics
  Communications, 185, 1246

\bibitem[{{Leconte} \& {Chabrier}(2013)}]{leconte}
{Leconte}, J. \& {Chabrier}, G. 2013, Nature Geoscience, 6, 347

\bibitem[{{Martins}(2004)}]{Martins}
{Martins}, R.~M. 2004, {Electronic Structure} (Cambridge University Press)

\bibitem[{{Mazevet} {et~al.}(2007){Mazevet}, {Lambert}, {Bottin}, {Z{\'e}rah},
  \& {Cl{\'e}rouin}}]{mazevet_2007}
{Mazevet}, S., {Lambert}, F., {Bottin}, F., {Z{\'e}rah}, G., \& {Cl{\'e}rouin},
  J. 2007, \pre, 75

\bibitem[{{Mazevet} {et~al.}(2019{\natexlab{a}}){Mazevet}, {Licari},
  {Chabrier}, \& {Potekhin}}]{mazevet_2019b}
{Mazevet}, S., {Licari}, A., {Chabrier}, G., \& {Potekhin}, A.~Y.
  2019{\natexlab{a}}, \aap, 621, A128

\bibitem[{{Mazevet} {et~al.}(2019{\natexlab{b}}){Mazevet}, {Licari}, \&
  {Soubiran}}]{mazevet_2019a}
{Mazevet}, S., {Licari}, A., \& {Soubiran}, F. 2019{\natexlab{b}}, submitted
  for publication

\bibitem[{{Mazevet} {et~al.}(2015){Mazevet}, {Tsuchiya}, {Taniuchi},
  {Benuzzi-Mounaix}, \& {Guyot}}]{mazevet_2015}
{Mazevet}, S., {Tsuchiya}, T., {Taniuchi}, T., {Benuzzi-Mounaix}, A., \&
  {Guyot}, F. 2015, \prb, 92, 014105

\bibitem[{{Miguel} {et~al.}(2016){Miguel}, {Guillot}, \& {Fayon}}]{Miguel_2016}
{Miguel}, Y., {Guillot}, T., \& {Fayon}, L. 2016, \aap, 596, A114

\bibitem[{{Militzer} \& {Hubbard}(2013)}]{Militzer_2013}
{Militzer}, B. \& {Hubbard}, W.~B. 2013, \apj, 774, 148

\bibitem[{{Morard} {et~al.}(2011){Morard}, {Bouchet}, {Valencia}, {Mazevet}, \&
  {Guyot}}]{morard_2011}
{Morard}, G., {Bouchet}, J., {Valencia}, D., {Mazevet}, S., \& {Guyot}, F.
  2011, High Energy Density Physics, 7, 141

\bibitem[{{Musella} {et~al.}(2019){Musella}, {Mazevet}, \&
  {Guyot}}]{musella_2019}
{Musella}, R., {Mazevet}, S., \& {Guyot}, F. 2019, \prb, 99, 064110

\bibitem[{{Nettelmann} {et~al.}(2012){Nettelmann}, {Becker}, {Holst}, \&
  {Redmer}}]{Nettelmann_2012}
{Nettelmann}, N., {Becker}, A., {Holst}, B., \& {Redmer}, R. 2012, \apj, 750,
  52

\bibitem[{{Nettelmann} {et~al.}(2013){Nettelmann}, {P{\"u}stow}, \&
  {Redmer}}]{Nettelmann_2013}
{Nettelmann}, N., {P{\"u}stow}, R., \& {Redmer}, R. 2013, \icarus, 225, 548

\bibitem[{{Poirier}(2004)}]{Poirier}
{Poirier}, J.~P. 2004, {Introduction to the physics of the earth's interior}
  (Cambridge University Press)

\bibitem[{{Pollack} {et~al.}(1996){Pollack}, {Hubickyj}, {Bodenheimer},
  {Lissauer}, {Podolak}, \& {Greenzweig}}]{Pollack_etal96}
{Pollack}, J.~B., {Hubickyj}, O., {Bodenheimer}, P., {et~al.} 1996, \icarus,
  124, 62

\bibitem[{{Soubiran} \& {Militzer}(2015)}]{Soubiran_2015}
{Soubiran}, F. \& {Militzer}, B. 2015, \apj, 806, 228

\bibitem[{{Stixrude}(2012)}]{stixrude_2012}
{Stixrude}, L. 2012, \prl, 108, 055505

\bibitem[{Umemoto \& Wentzcovitch(2011)}]{umemoto_2011}
Umemoto, K. \& Wentzcovitch, R.~M. 2011, Earth and Planetary Science Letters,
  311, 225

\bibitem[{{Umemoto} {et~al.}(2006){Umemoto}, {Wentzcovitch}, \&
  {Allen}}]{umemoto_2006}
{Umemoto}, K., {Wentzcovitch}, R.~M., \& {Allen}, P.~B. 2006, Science, 311, 983

\bibitem[{{von Zahn} {et~al.}(1998){von Zahn}, {Hunten}, \&
  {Lehmacher}}]{galileo}
{von Zahn}, U., {Hunten}, D.~M., \& {Lehmacher}, G. 1998, Journal of
  Geophysical Research, 103, 22815

\bibitem[{{Wahl} {et~al.}(2017{\natexlab{a}}){Wahl}, {Hubbard}, {Militzer},
  {Guillot}, {Miguel}, {Movshovitz}, {Kaspi}, {Helled}, {Reese}, {Galanti},
  {Levin}, {Connerney}, \& {Bolton}}]{Wahl_etal17}
{Wahl}, S.~M., {Hubbard}, W.~B., {Militzer}, B., {et~al.} 2017{\natexlab{a}},
  Geophys.\ Research Lett., 44, 4649

\bibitem[{{Wahl} {et~al.}(2017{\natexlab{b}}){Wahl}, {Hubbard}, {Militzer},
  {Guillot}, {Miguel}, {Movshovitz}, {Kaspi}, {Helled}, {Reese}, {Galanti},
  {Levin}, {Connerney}, \& {Bolton}}]{Wahl_2017}
{Wahl}, S.~M., {Hubbard}, W.~B., {Militzer}, B., {et~al.} 2017{\natexlab{b}},
  Geophysical Research Letters, 44, 4649

\bibitem[{{Wilson} \& {Militzer}(2012)}]{Wilson_2012}
{Wilson}, H.~F. \& {Militzer}, B. 2012, \apj, 745, 54

\bibitem[{{Wilson} {et~al.}(2013){Wilson}, {Wong}, \& {Militzer}}]{wilson_2013}
{Wilson}, H.~F., {Wong}, M.~L., \& {Militzer}, B. 2013, \prl, 110

\bibitem[{Wu {et~al.}(2013)Wu, Ji, Wang, Nguyen, Zhao, Umemoto, Wentzcovitch,
  \& Ho}]{Wu_2013}
Wu, S.~Q., Ji, M., Wang, C.~Z., {et~al.} 2013, Journal of Physics: Condensed
  Matter, 26, 035402

\bibitem[{{Zhou} \& {Lin}(2007)}]{Lin_2007}
{Zhou}, J.-L. \& {Lin}, D. N.~C. 2007, \apj, 666, 447

\end{thebibliography}
   \appendix
   \section{Method}
   We carried out the {\sl \textup{ab initio}} molecular dynamics simulations using the ABINIT \citep{2009CoPhC.180.2582G}
   electronic structure package. {\sl \textup{Ab initio}} molecular dynamics simulations consist of treating the electrons quantum mechanically using the finite temperature
   density functional theory (DFT) while propagating the ions classically on the resulting Born--Oppenheimer surface
   by solving the Newton equations \citep{Martins}. The molecular dynamics runs were performed using the finite temperature
   formulation of DFT as laid out by Mermin \citep{Martins}. The equations of motion for the ions were integrated using
   the iso-kinetics ensemble \citep{tidsley}. For each simulation, this consists of keeping the number of particles
   as well as the volume of the simulation cell fixed while rescaling the atom velocities at each time step to keep the
   temperature constant. While it is well documented that this ensemble does not formally correspond to the canonical
   ensemble, it is used here in a situation where the property calculated, the melting temperature, is not sensitive
   to the use of a more refined thermostat \citep{tidsley}. We typically used a time step that varied from $0.1$ to $1~$fs
   for the various systems we investigated.

   \subsection{MgSiO$_3$}
   We obtained
   the EOS points by performing simulations at the $\Gamma$-point and using 164 atoms in the simulation cell. To reach the
   high-densities corresponding to pressures that are representative of the cores of giants and ice giants, we developed
   customized projector augmented wave (PAW) pseudo-potentials with small pseudization radius and several shells as
   valence shells to ensure non-overlapping PAW spheres. This also allowed us to include the potential effect of core
   electrons at high pressures.

   This led to pseudo-potentials that only consider the $1s^2$ shell as a core shell and use a cutoff radius of
   r$_{paw}=1.1~$ a$_B$ for the Mg and Si atoms and r$_{paw}=1.0~$a$_B$ for the O atoms. This small cutoff radius led
   to large plane-wave basis sets with a cutoff energy of $E_{cut}=1000~$eV to converge the physical
   properties such as the pressure and the internal energy to less than 1\%. The accuracy of the pseudo-potentials
   produced were tested for the $T=O$K EOS (the cold curve) against all electron calculations for
   the individual elements to ensure that now spurious effects were introduced in the pseudization
   procedure \citep{jollet}.

   The molecular dynamics simulations were performed after first relaxing the non-cubic
   PPV cell along the three directions, $(a,b,c)$. This was done to obtain the pressure dependence of the $a/b$ and $c/a$
   ratio. We found that the $b/a$ ratio decreases by 15\% when the pressure increases to $30$ Mbar, while the $c/a$
   ratio remains almost constant.

   \subsection{H$_2$O}
   The computational details regarding the pseudo-potentials we used and the convergence criteria can be found in \cite{mazevet_2019a}.
   The direct melting simulations were performed using 54 and 108 atoms in the simulation cells for the BCC and FCC super-ionic structure, respectively.
   The two-phase calculations in the BCC phase were performed using 108 atoms in the simulation cell.   
   \subsection{Fe}
   The high-pressure EOS points were calculated by creating short range
   pseudo-potentials with a cutoff radius of $r_{paw}= 1.4$a$_B$ and considering the $3s3p4s4p$ states as valence states. The two-phase simulations were performed using
   108 and 216 atoms in the simulation cell for the BCC and FCC structures, respectively.
   \subsection{Thomas--Fermi}
   This high-density method consists of approximating the kinetic term by a functional of the
   density; see \cite{Martins}. As demonstrated previously \citep{mazevet_2007}, the Thomas--Fermi approximation is the natural DFT
   extension when the pressures obtained by each method coincide. In this method, the pseudo-potential is recalculated at each pressure-temperature condition that is
   simulated. The pseudo-potential is thus not transferable. The two-phase simulations were performed using 256 and 216 atoms for the BCC and FCC structures, respectively.

  \subsection{Parameterization of the calculated high-pressure melting temperatures}

   For convenient use in planetary modeling, we fit the ab initio\textup{} high-pressure melting temperatures we obtained for each element using a Simon law \cite{Poirier}. This empirical law relates the melting temperature $T_m$ to the pressure using the relation
 \begin{equation}
  T_M=T_{ref}\left(\frac{P_M-P_{ref}}{a}+1\right)^{(1/c)},
\end{equation}
where $T_{ref}$ and $P_{ref}$ are the temperature and pressure adjusted to match the phase boundary, while $a$ and $c$ are two adjustable parameters. 
   \begin{table}[ht]
\begin{center}
  \begin{tabular}{|c|c|c|c|c|}
    \hline
 Element/Phase & $T_{ref}$(K)& $P_{ref}$(Mbar)&$a$(Mbar)&$c$\\
\hline 
Fe-HCP &1800  & 0 & 0.31  & 1.99 \\  
\hline
Fe-FCC &17360 &29.34  & 5.49 &3.17\\
\hline
H2O & 4500& 1.21& 0.47& 4.75\\
\hline
MgSiO$_3$ &6277& 1.472 &0.174 & 4.39\\
\hline
MgO & 9200& 2.52&1.26 & 3.30\\
\hline
SiO$_2$ & 3300& 0.2& 0.29& 3.24\\
\hline
  \end{tabular}
  \caption{Coefficients of the Simon-Glatzel parameterization fitting the {\sl \textup{ab initio}} simulation results. The Fe-HCP values are from \cite{bouchet}, MgO is from \cite{musella_2019}, and SiO$_2$ from \cite{mazevet_2015}.}.
  \label{tab2}
\end{center}
\end{table}
\end{document}